\documentclass{pic2012}
\def\runauthor{Antonino Sergi}
\def\shorttitle{Recent results from Kaon Physics}
\def \kpnn {K^{+} \rightarrow \pi^{+}\nu\overline{\nu}}
\def \kpnnn {K \rightarrow \pi \nu\overline{\nu}}
\def \klnn {K_L \rightarrow \pi^{0}\nu\overline{\nu}}
\newcommand{\alert}[1] {#1}
\begin{document}

\title{Recent results from Kaon Physics}

\author{Antonino Sergi}

\address{CERN, Geneva, Switzerland}

\maketitle

\abstracts{A short review of recent results and future prospects in kaon physics is presented.
Recent measurements performed at the NA48, NA62, KLOE and KTeV experiments
on CP and Lepton Flavour violation and rare decays will be summarised,
together with measurements of CKM elements and Chiral Perturbation Theory tests.
}

\section{Introduction}
$K$-mesons (kaons) were discovered in 1947 in cosmic rays, and produced in laboratory few years later; they were the first
particles not fitting with the light flavour scheme and brought to the introduction of a new quantum number,
called Strangeness, violated only by weak interactions. Experiments showed kaons to have an unprecedented behavior;
new assumptions were made to explain their phenomenology, especially related to neutral kaons, in terms of new properites.
In particular the analysis of their behavior under $CP$ transformation had major contributions to establish the basis
of the modern Standard Model (SM) of particle physics. The discovery of $CP$ violation by $K_{1}$ ($\frac{1}{\sqrt{2}} (K^{0} + \overline{K}^{0})$)
decaying in $\pi^{+}\pi^{-}\pi^{0}$ 
and $K_{2}$ ($\frac{1}{\sqrt{2}} (K^{0} - \overline{K}^{0})$) decaying in $\pi^{+}\pi^{-}$
showed that the mass eigenstates are $K_{S} = K_{1} + \epsilon K_{2}$
and $K_{L} = K_{2} + \epsilon K_{1}$, where $\epsilon$ is the indirect $CP$ violation (mixing) parameter.
\subsection{$CP$}
$CP$ violation in the Standard Model is described by a single complex phase in the Cabibbo-Kobayashi-Maskawa (CKM) mass-eigenstate mixing matrix.
The Kobayashi-Maskawa mechanism \cite{KM1,KM2,KM3}, accommodating $CP$ violation in the electroweak theory, predicted the third generation of quarks.
There is a set of parameters used to describe $CP$ violation phenomenology and connect it with the theory: $\epsilon~O(10^{-3})$, already mentioned,
$\epsilon'~O(10^{-6})$, giving the direct $CP$ violation (decay), and
$\eta_{+-}  = \frac{K_{L} \rightarrow \pi^{+}\pi^{-}}{K_{S} \rightarrow \pi^{+}\pi^{-}} = \epsilon + \epsilon'$,
$\eta_{00}  = \frac{K_{L} \rightarrow \pi^{0}\pi^{0}}{K_{S} \rightarrow \pi^{0}\pi^{0}} = \epsilon - 2 \epsilon'$,
$\Delta\phi = \phi_{00} - \phi_{+-} = -3 Im(\frac{\epsilon'}{\epsilon})$.
After the unexpected discovery of the $CP$-violating $K_{L} \rightarrow \pi^{+}\pi^{-}$ decay in 1964 \cite{CF}, $\epsilon$ was measured \cite{Cronin,Fitch}
exploiting the interference between semi-leptonic decays
($2 Re(\epsilon) = \frac{K_{L} \rightarrow \pi^{-}l^{+}\nu ~ - ~ K_{L} \rightarrow \pi^{+}l^{-}\overline\nu}
                                {K_{L} \rightarrow \pi^{-}l^{+}\nu ~ + ~ K_{L} \rightarrow \pi^{+}l^{-}\overline\nu}$),
instead of measuring directly $\eta_{+-}$ or $\eta_{00}$.
The measurement of $\epsilon'$ was achieved at the end of the century, exploiting the double ratio 
$|\frac{\eta_{00}}{\eta_{+-}}|^{2} = 1 - 6 Re(\frac{\epsilon'}{\epsilon})$ to increase the sensitivity of the experiments.

\subsection{Chiral Perturbation Theory}
Most kaon decays are governed by long distance physics, involving non perturbative QCD; to study their properties an effective field theory in
terms of QCD Goldstone bosons has been developed, namely Chiral Perturbation Theory (ChPT). It is an expansion in powers of momenta 
and quark masses over $\Lambda_\chi \approx 1$ GeV and provides a theoretical framework both for (semi)leptonic and nonleptonic decays, including radiative decays,
by means of a pseudoscalar-octet and electroweak operators. A set of Low Energy Constants (LECs) has to be extracted from experiments, by measuring Form Factors (FF),
to be able to make predictions and compare with other experimental results.

\section{$CP$ violation and CKM matrix}
\subsection{$CP$ violation}
Measuring $\epsilon'/\epsilon$ required to measure all the 4 involved decays simultaneously to exploit cancellation of systematics in the double ratio
$|\frac{\eta_{00}}{\eta_{+-}}|$.
NA48 at CERN and KTeV at FNAL were designed to do so: intense $K_{L}$ beams at high momentum (for $K_{L} \rightarrow \pi^{0}\pi^{0}$) with decay regions 
$\approx 100 m$ for both experiments, while the production of $K_{S}$ was by means of a regenerator (KTeV) or a second target close to the decay region (NA48).
The final result for $Re(\frac{\epsilon'}{\epsilon})$ was $(2.071 \pm 0.148_{stat} \pm 0.239_{syst}) 10^{-3} = (2.07 \pm 0.28) 10^{-3}$ for KTeV \cite{EpsilonPrimeKTeV}
and $(1.47 \pm 0.14_{stat} \pm 0.09_{stat/syst} \pm 0.15_{syst}) 10^{-3} = (1.47 \pm 0.22) 10^{-3}$ for NA48 \cite{EpsilonPrimeNA48}. Unfortunately the poor precision
of the theoretical prediction \cite{EpsilonPrimelQCD}, based on lattice QCD (lQCD), prevented to fully exploit this measurement, but there is currently a new approach
which uses the experimental value as input for lQCD calculations \cite{EpsilonPrimelQCD2}.

Another tool to explore $CP$ violation is the study of $K_S \rightarrow \pi^0 \pi^0 \pi^0$ decay; 
$\eta_{000}  = \frac{K_{L} \rightarrow 3\pi^{0}}{K_{S} \rightarrow 3\pi^{0}} = \epsilon + \epsilon_{000}'$ can be defined, with 
$\epsilon_{000}'= -2\epsilon'$ to lowest order in ChPT. The Standard Model prediction $BR(K_{S} \rightarrow 3\pi^{0}) = 1.9 \times 10^{-9}$ has been
out of reach up to now, but several upper limits were imposed in the past 10 years respectively by SND, NA48 and KLOE, $1.4 \times 10^{-5}$ in 1999, 
$7.4 \times 10^{-7}$ in 2004, $1.2 \times 10^{-7}$ in 2005 and $2.7 \times 10^{-8}$ in 2012, being the last 2 by KLOE. A first observation should
be feasible in KLOE-2, because of an improved inner tracker and a better photon coverage near the interaction point.

$CP$ violation can be studied also in the decay of charged kaons by defining charge asymmetries:
given $\Gamma(K^{\pm} \rightarrow \pi^{\pm} \pi \pi) \propto 1 + g \cdot u + h \cdot u^2 + k \cdot v^2$, the asymmetry $A_g = \frac{g^+ - g^-}{g^+ + g^-}$
represents $CP$ violation in decay, with an expectation of $O(10^{-5} - 10^{-6})$; in the past few years NA48/2 has produced several results:
$A_g(K^{\pm} \rightarrow \pi^{\pm} \pi^+ \pi^-)$=$ (-1.5 \pm 2.2) 10^{-4}$,
$A_g(K^{\pm} \rightarrow \pi^{\pm} \pi^0 \pi^0)$=$ ( 1.8 \pm 1.8) 10^{-4}$,
$A_g(K^{\pm} \rightarrow \pi^{\pm} \pi^0 \gamma)$=$( 0.0 \pm 1.2) 10^{-3}$,
$A_g(K^{\pm} \rightarrow \pi^{\pm} e^+ e^-)$=$(-2.2 \pm 1.6)10^{-2}$,
$A_g(K^{\pm} \rightarrow \pi^{\pm} \mu^+ \mu^-)$=$ ( 1.2 \pm 2.3) 10^{-2}$.
%
\subsection{$CPT$}
\begin{figure}[!htbp]
\begin{center}
\begin{minipage}[!htbp]{0.45\textwidth}
        \centering
        \includegraphics[width=\textwidth,height=0.4\textwidth]{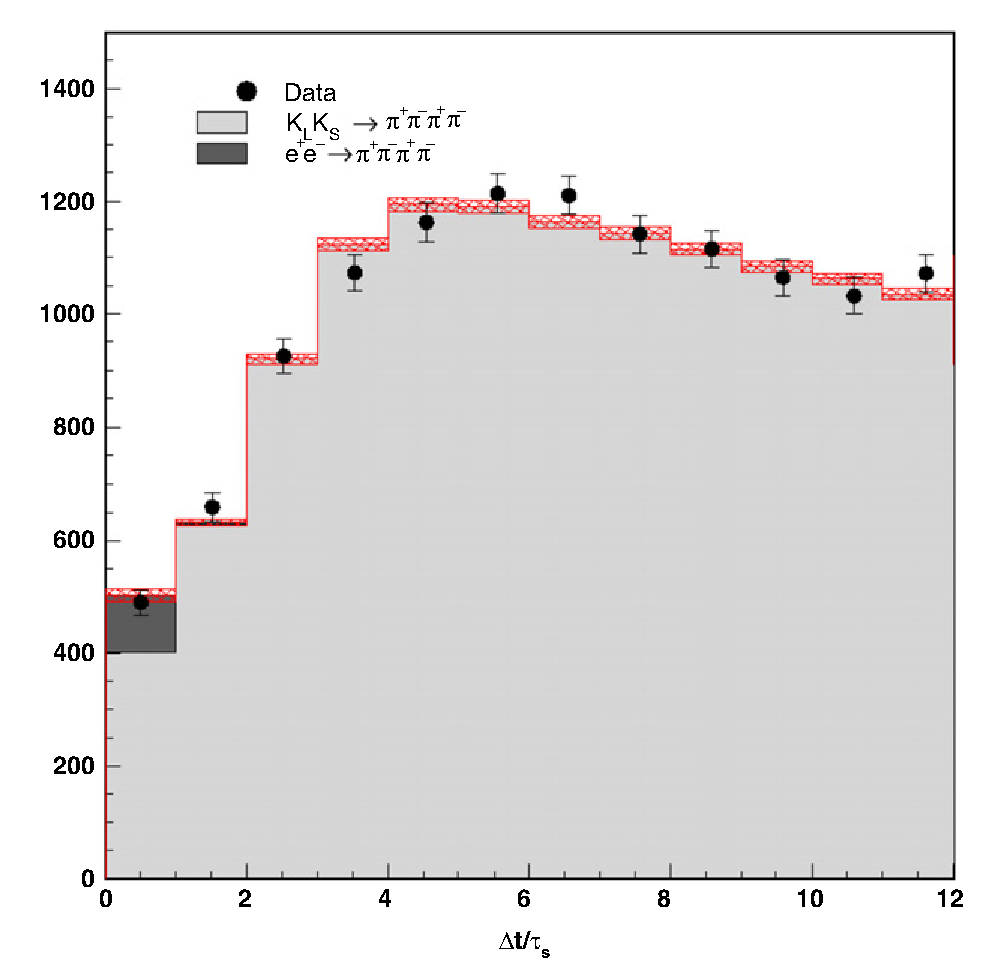}
        \caption{$I(\Delta t)$ to measure $\zeta_{SL}$}
        \label{fig:KLOECPTQM}
\end{minipage}
\begin{minipage}[!htbp]{0.45\textwidth}
        \centering
        \includegraphics[width=\textwidth,height=0.4\textwidth]{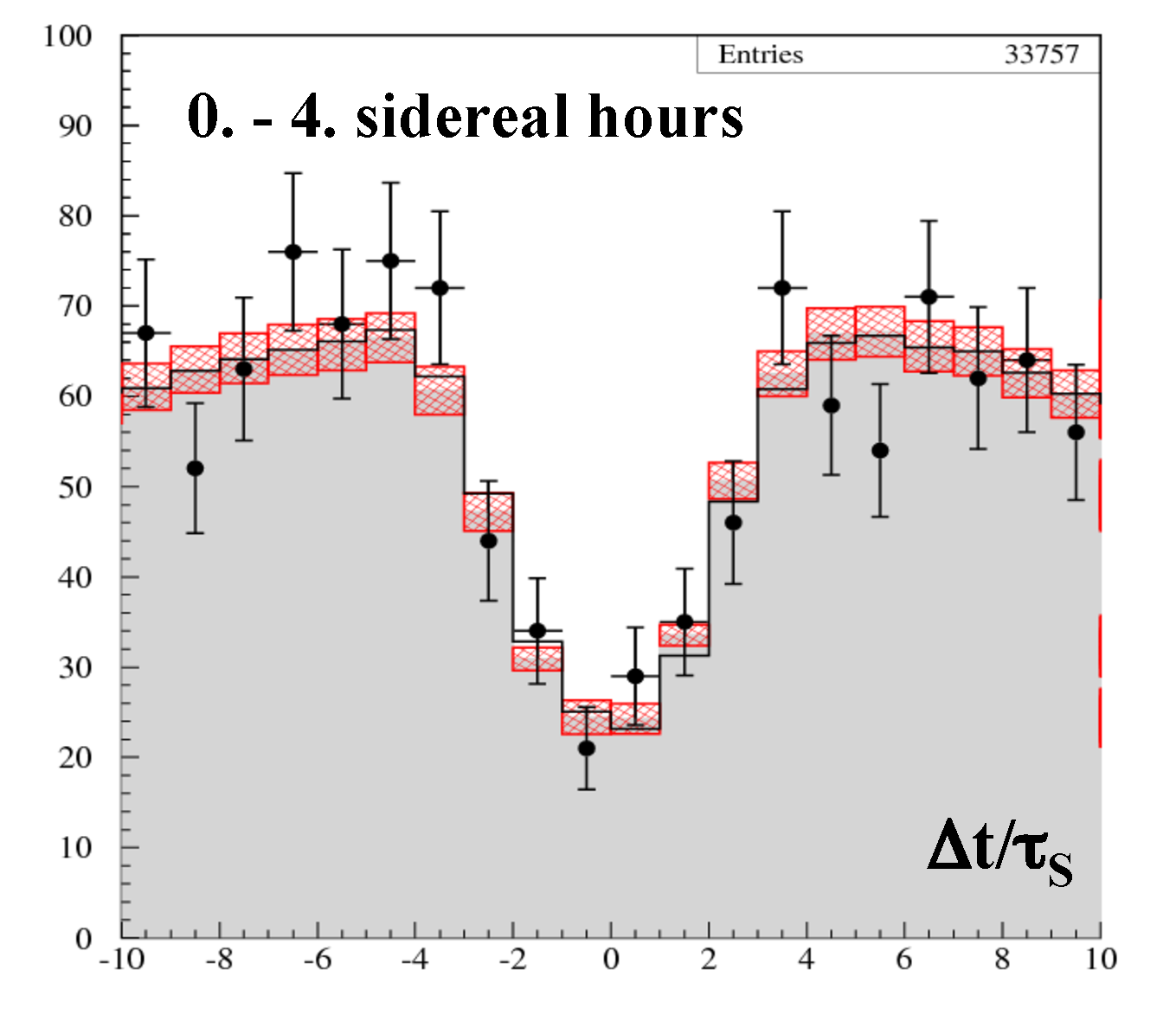}
        \caption{$I(\Delta t)$ to measure $\Delta \vec a$}
        \label{fig:KLOECPTSMEMeas}
\end{minipage}
\end{center}
\end{figure}

Test of $CPT$ invariance and quantum mechanics can also be performed by studying kaon decays.
In the $CP$-violating process $\phi \rightarrow K_S K_L \rightarrow \pi^+\pi^- \pi^+\pi^-$ an intensity 
$I(\Delta t) \propto e^{-\Gamma_L \Delta t} + e^{-\Gamma_S \Delta t} - 2 (1 - \zeta_{SL}) e^{-\frac{\Gamma_L + \Gamma_S}{2} \Delta t} cos(\Delta m\Delta t)$,
can be defined, with $\Delta m = m_{K_L} - m_{K_S}$, $\Delta t$ the decay time difference and $\zeta_{SL}$ as decoherence parameter; for $\Delta t \rightarrow 0$,
$I(\Delta t)\rightarrow 2\zeta_{SL}\left(1 - \frac{\Gamma_L + \Gamma_S}{2} \Delta t \right)$ and $\zeta_{SL}$ can be extracted.
$\zeta_{SL} = 0.018 \pm 0.040_{stat} \pm 0.007_{syst}$ was obtained by the KLOE experiment (fig. \ref{fig:KLOECPTQM}) \cite{KLOECPT1}.

A different approach can give information about $CPT$ and Lorentz invariance; there is a Standard Model Extension (SME) consisting in a phenomenological 
effective model providing a framework for CPT and Lorentz violation \cite{SME1,SME2}. If we define $\epsilon_{S,L} = \epsilon \pm \delta$, with
$\delta = i \sin \phi_{SW} e^{i\phi_{SW}} \gamma_K (\Delta a_0 - \vec \beta_K \cdot \Delta \vec a)/\Delta m$,
$\Delta a_0$ and $\Delta \vec a$ are four parameters associated to SME lagrangian terms and related to CPT and Lorentz violation. Exploiting interferometry, 
$I(\Delta t) \propto |\eta_1|^2 e^{-\Gamma_L \Delta t} + |\eta_2|^2 e^{-\Gamma_S \Delta t} - 2 |\eta_1||\eta_2| e^{-\frac{\Gamma_L + \Gamma_S}{2} \Delta t} cos(\Delta m\Delta t)$ can be defined, where $\eta_1^{+-} = \epsilon (1 - \delta(\vec p,t))$ and $\eta_2^{+-} = \epsilon (1 - \delta(- \vec p,t))$. The measurement
of $Im(\delta)$ can be done at small $\Delta t$ while $Re(\delta)$ at large $\Delta t$. With this technique KLOE reached a preliminary result (1 fb$^{-1}$ integrated
luminosity) for
$\Delta a_x$, $\Delta a_y$ and $\Delta a_z$, $(-6.3 \pm 6.0) \times 10^{-18}$ GeV, $( 2.8 \pm 5.8) \times 10^{-18}$ GeV and $( 2.4 \pm 9.7) \times 10^{-18}$ GeV
respectively (fig. \ref{fig:KLOECPTSMEMeas}); they should be compared with a previous result from KTeV 
($\Delta a_x$, $\Delta a_y < 9.2 \times 10^{-22}$ GeV), but it is the first
measurement of $\Delta a_z$.
%
\subsection{$V_{us}$}
The semileptonic decays usually called $K_{l3} (K \rightarrow \pi^0 e \nu_e$, $K \rightarrow \pi^0 \mu \nu_\mu)$ can be used to extract an effective measurement
of $|V_{us}|$; the decay amplitude $\Gamma(K_{l3(\gamma)})$ can be parametrized as 
$\frac{m_K^5 G_F^2}{192 \pi^3} C_K^2 S_{EW} |V_{us}|^2 |f_+(0)|^2 I_K^l (1 + 2 \delta_{SU(2)}^l + 2 \delta_{EM}^l)$, where $C_K^2 = 1$ for $K^0$, 
$=1/2$ for $K^{\pm}$ and  $S_{EW} = 1.0232$ is the short distance EW correction. $\Gamma(K_{l3(\gamma)})$ and $I_K^l$ (form factors integral) can be
extracted from experiments, while $f_+(0)$ (hadronic matrix element at $q^2 = 0$), $\delta_{SU(2)}^l$, $\delta_{EM}^l$ (SU(2) breaking and long distance EM corrections)
are the results of theoretical calculations, being the first, $f_+(0)$, entirely a result of lQCD. Its value and uncertainty are as crucial as the
experimental result for the extraction of $|V_{us}|$. The FlaviaNet collaboration, combining all the available measurements and calculations, in 2010 
reached a determination of $|V_{us}| = 0.2254 \pm 0.0013$; another result is $\Delta_{CKM} = -0.0001 \pm 0.0006$, with 
$\Delta_{CKM} \equiv |V_{ud}|^2 + |V_{us}|^2 + |V_{ub}|^2 - 1$, as test of the unitarity of the CKM matrix.
\section{Chiral Perturbation Theory}
The same decays ($K_{l3}$) are useful, by measuring their FF, to obtain some of the LECs of ChPT. The matrix element can be written as 
$M = \frac{G_F}{2}|V_{us}|(f_+(t)(P_K + P_\pi)^\mu \overline{u}_l \gamma_\mu (1 + \gamma_5)u_\nu + f_-(t) m_l \overline{u}_l (1 + \gamma_5)u_\nu)$, $t=q^2$;
the scalar FF $f_0(t)$ can be expressed as linear combiation of vector FF: $f_0(t) = f_+(t) + \frac{t}{m_K^2 - m_\pi^2} f_-(t)$, where $f_+(0)$ is not measurable 
but $\overline{f}_+(t) = \frac{f_+(t)}{f_+(0)}$ and $\overline{f}_0(t) = \frac{f_0(t)}{f_+(0)}$ are experimentally accessible.
Two parametrizations are used: the first one ($\overline{f}_{+,0}(t) = \frac{m_{V,S}^2}{m_{V,S}^2 - t}$), usually known as Pole, assumes the exchange 
of a vector($1^-)$ or scalar ($0^+$) resonances ($m_{V,S}$), while the second, a linear ($\overline{f}_{+,0}(t) = 1 + \lambda_{+,0} \frac{t}{m_{\pi}^2}$) 
or quadratic ($\overline{f}_{+,0}(t) = 1 + \lambda_{+,0}' \frac{t}{m_{\pi}^2} + \lambda_{+,0}'' \left(\frac{t}{m_{\pi}^2}\right)^2$) expansion has no 
physical meaning.
NA48/2 has preliminary results from $K \rightarrow \pi^0 e \nu_e$, $K \rightarrow \pi^0 \mu \nu_\mu$, which can be taken separately for the decays with an electron 
or a muon. For the quadratic expansion ($\times 10^{-3}$) $\lambda_+' = 26.3 \pm 3.0_{stat} \pm 2.2_{syst}$, $\lambda_+''=1.2 \pm 1.1_{stat} \pm 1.1_{syst}$ and 
$\lambda_0'=15.7 \pm 1.4_{stat} \pm 1.0_{syst}$ for $K_{\mu 3}$ while $\lambda_+' = 27.2 \pm 0.7_{stat} \pm 1.1_{syst}$ and 
$\lambda_+''= 0.7 \pm 0.3_{stat} \pm 0.4_{syst}$ for $K_{e3}$. The Pole gives $m_V = (873 \pm 8_{stat} \pm 9_{syst})$ MeV/$c^{2}$ and 
$m_S=(1183 \pm 31_{stat} \pm 16_{syst})$ MeV/$c^{2}$ for $K_{\mu 3}$ while $m_V = (879 \pm 3_{stat} \pm 7_{syst})$ MeV/$c^{2}$ for $K_{e3}$. The combined result is
$\lambda_+' = (26.91 \pm 1.11)10^{-3}$, $\lambda_+''=(0.81 \pm 0.46)10^{-3}$, $\lambda_0'=(16.23 \pm 0.95)10^{-3}$, $m_V = (877 \pm 6)$ MeV/$c^{2}$,
$m_S=(1176 \pm 31)$ MeV/$c^{2}$.

Results for $K_{e3}$ and $K_{\mu3}$ from NA48/2 are in good agreement and, given their high precision, they are competitive with other measurements, especially
if the smallest error in the combined result is considered.
\begin{figure}[!htbp]
\begin{center}
\begin{minipage}[!htbp]{0.45\textwidth}
        \centering
        \includegraphics[width=\textwidth,height=0.5\textwidth]{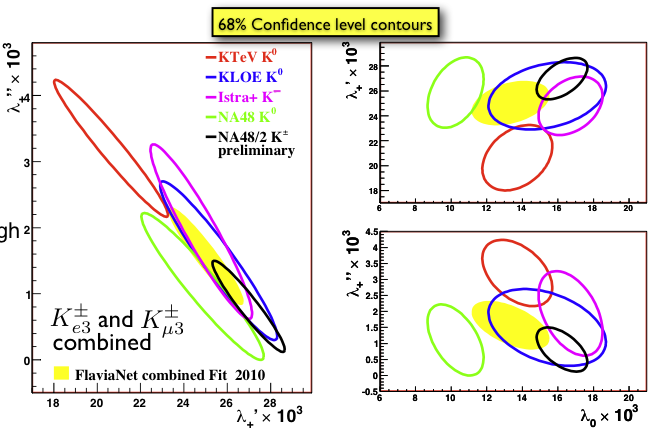}
        \caption{$K_{e3}$ and $K_{\mu3}$ combined}
        \label{fig:Kl3C2comb}
\end{minipage}
\begin{minipage}[!htbp]{0.45\textwidth}
        \centering
        \includegraphics[width=\textwidth,height=0.5\textwidth]{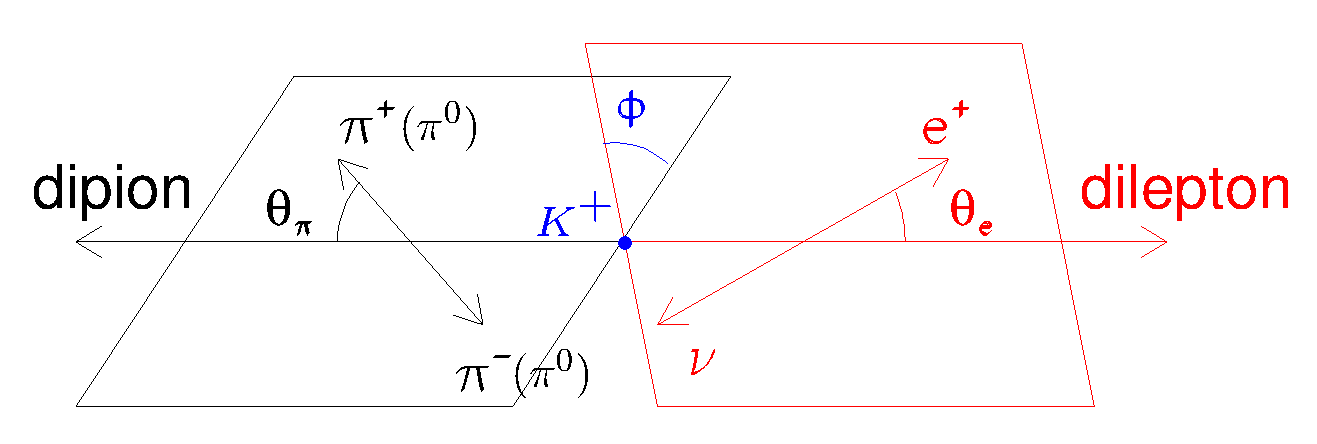}
        \caption{$K_{e4}$ decay geometry}
        \label{fig:Ke4Geo}
\end{minipage}
\end{center}
\end{figure}

An important family of kaon decays is the one named $K_{e4}$ ($K \rightarrow \pi^+ \pi^- e \nu_e$, called $K_{e4}(+-)$ and 
$K \rightarrow \pi^0 \pi^0 e \nu_e$, called $K_{e4}(00)$)
The geometry of the decay, shown in fig. \ref{fig:Ke4Geo}, allows to describe it by five kinematic variables \footnote{Cabibbo-Maksymowicz 1965}:
$s_\pi=M_{\pi\pi}^2$, $s_e=M_{e\nu}^2$, $cos \theta_\pi$, $cos \theta_e$ and $\phi$.

$K_{e4}$ FF can be expressed by a partial wave expansion, limited to S and P waves \footnote{Pais-Treiman (1968) + Watson theorem (T invariance)}; it consists
in 2 axial FF ($F = F_s e^{i\delta_s} + F_p e^{i\delta_p} cos \theta_\pi$ and $G = G_p e^{i\delta_p}$) and 1 vector FF ($H = H_p e^{i\delta_p}$), with real fit
parameters ($F_s$, $F_p$ , $G_p$ , $H_p$, $\delta=\delta_s - \delta_p$ for the charged decays, and $F_s$ only for the neutral ones). The $q^2$ dependence 
can be studied from FF fitted in $q^2$ bins \cite{Ke4FFPars} using
$F_s^2 = f_s^2 \left[1 + \frac{f_s'}{f_s} q^2 + \frac{f_s''}{f_s} q^4 + \frac{f_e'}{f_s} \frac{M_{e\nu}^2}{4m_\pi^2}\right]$ and 
$\frac{G_p}{f_s}=\frac{g_p}{f_s} + \frac{g_p'}{f_s} q^2$, $F_p = f_p$, $H_p = h_p$, with $q^2 = \left[\frac{M_{\pi\pi}^2}{4m_\pi^2} - 1\right]$.

NA48/2 total statistics (2003+2004) allowed to measure $K_{e4}(+-)$ the relative FF \cite{NA482Ke4FFr}
\begin{table}[!htbp]
\begin{center}
\begin{minipage}[!htbp]{0.45\textwidth}
        \centering
\begin{tabular}{|c|r c c|}
	\hline
				& value	& stat	& syst\\
	\hline
	$\frac{f_s'}{f_s}$	& 0.152	& $\pm0.007$	& $\pm0.005$\\
	$\frac{f_s''}{f_s}$	&-0.073	& $\pm0.007$	& $\pm0.006$\\
	$\frac{f_e'}{f_s}$	& 0.068	& $\pm0.006$	& $\pm0.007$\\
	\hline
	$\frac{f_p}{f_s}$	&-0.048	& $\pm0.003$	& $\pm0.004$\\
	\hline
	$\frac{g_p}{f_s}$	& 0.868	& $\pm0.010$	& $\pm0.010$\\
	$\frac{g'_p}{f_s}$	& 0.089	& $\pm0.017$	& $\pm0.013$\\
	\hline
	$\frac{h_p}{f_s}$	&-0.398	& $\pm0.015$	& $\pm0.008$\\
	\hline
\end{tabular}
        \caption{$K_{e4}(+-)$ the relative FF}
        \label{tab:Ke4FFr}
\end{minipage}
\begin{minipage}[!htbp]{0.45\textwidth}
        \centering
\begin{tabular}{l c}
	\hline
	Relative Systematic Uncertainty & (\%)\\
	\hline
	Acceptance, beam geom. 		& 0.18\\
	Muon vetoing 			& 0.16\\
	Accidental activity 		& 0.21\\
	Particle ID			& 0.09\\
	Background			& 0.07\\
	Radiative effects		& 0.08\\
	Trigger efficiency		& 0.11\\
	Simulation statistics		& 0.05\\
	\hline
	Total systematics		& 0.37\\
	External error [$BR(K_{3\pi})$]	& 0.72\\
	\hline
\end{tabular}
        \caption{$K_{e4}(+-)$ uncertainties}
        \label{tab:Ke4BRc}
\end{minipage}
\end{center}
\end{table}

Once the relative FF are available, $K_{e4}(+-)$ branching fraction is needed to obtain the absolute FF; this analysis \cite{NA482Ke4BR}. 
uses $K^\pm \rightarrow \pi^\pm \pi^+ \pi^-$ decays as normalization ($BR = (5.59 \pm 0.04)\%$), 
giving $(1.11 \times 10^6)$ signal events, background 0.95\% of $K_{e4}$ and a normalization statistics of $(1.9 \times 10^9)$ events,
with a signal and normalization acceptance of 18.19\% and 23.97\% and trigger efficiency 98.5\% and 97.7\% respectively. 
Results, to be compared with the current world average $(4.09 \pm 0.10) \times 10^{-5}$ \cite{PDG2012}, are
$BR(K_{e4}^+) = (4.255 \pm 0.008) \times 10^{-5}$, $BR(K_{e4}^-) = (4.261 \pm 0.011) \times 10^{-5}$ (never measured before), and, combining them, 
$BR[K_{e4}^\pm(+-)] = (4.257 \pm 0.004_{stat} \pm 0.016_{syst} \pm 0.031_{ext}) \times 10^{-5}$. 

Using $BR[K_{e4}^\pm(+-)]$ as an overall form factor normalization, $K_{e4}(+-)$ absolute FF (NA48/2) can be calculated \cite{NA482Ke4BR}. 
\begin{table}[!htbp]
\begin{center}
\begin{minipage}[!htbp]{0.5\textwidth}
        \centering
\begin{tabular}{l c}
	\hline
	Relative Systematic Uncertainty & (\%)\\
	\hline
	Background			& 0.35\\
	Simulation statistics		& 0.12\\
	Form factor dependence		& 0.20\\
	Radiative effects		& 0.23\\
	Trigger efficiency		& 0.80\\
	Particle ID			& 0.10\\
	Beam geometry			& 0.10\\
	\hline
	Total systematics		& 0.94\\
	External error [$BR(K_{3\pi})$]	& 1.25\\
	\hline
\end{tabular}
        \caption{$K_{e4}(00)$ uncertainties}
        \label{tab:Ke4BRn}
\end{minipage}
\begin{minipage}[!htbp]{0.4\textwidth}
        \centering
        \includegraphics[width=\textwidth]{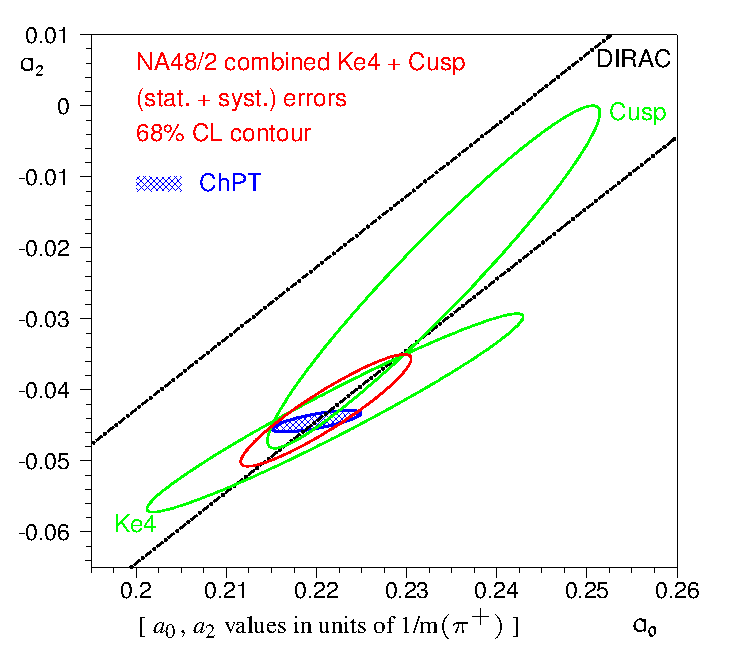}
        \caption{$\pi\pi$ scattering lengths $a_0$ , $a_2$}
        \label{fig:Ke4a0a2}
\end{minipage}
\end{center}
\end{table}

NA48/2 has also a preliminary result for $K_{e4}(00)$ branching fraction; the analysis, still in progress,
uses $K^\pm \rightarrow \pi^\pm \pi^0 \pi^0$  decays as normalization ($BR = (1.761 \pm 0.022)\%$), 
giving $(4.49 \times 10^4)$ signal events, background 1.3\% of $K_{e4}$ and a normalization statistics of $(71 \times 10^6)$ events,
with a signal and normalization acceptance of 1.77\% and 4.11\% and trigger efficiency between 92\% and 98\% respectively. 

The preliminary result, to be compared with the current world average $(2.2 \pm 0.4) \times 10^{-5}$ \cite{PDG2012}, is
$BR[K_{e4}^\pm(00)] = (2.595 \pm 0.012_{stat} \pm 0.024_{syst} \pm 0.032_{ext}) \times 10^{-5}$

From $K_{e4}(+-)$ decay another physical information can be extracted: the $\pi\pi$ scattering lengths.
The S-wave $\pi\pi$ scattering lengths $a_0$ and $a_2$ (I = 0 and I = 2) are precisely predicted by ChPT \cite{ChPTA0A2,ChPTA0A2b}. NA48/2 performed two 
statistically independent measurements: one from the phase shift $\delta(M_{\pi\pi})=\delta_s - \delta_p$ in $K_{e4}$ decay \cite{NA482Ke4FFr} and
the second one from the “cusp” in $M_{\pi^0\pi^0}$ in $K^\pm \rightarrow \pi^\pm \pi^0 \pi^0$ decay \cite{NA482cusp}. These two measurements exibit
completely different systematics (electron misidentification and background vs. calorimetry and trigger), different theoretical inputs (Roy equations
and isospin breaking correction vs. rescattering in final state and ChPT expansion), and yet a large overlap in the $a_0$, $a_2$ plane and an impressive 
agreement with ChPT.

\subsection{Radiative}
Radiative decays are characterized by the presence of real photons in the final state, which can be produced by leptonic, semileptonic or non-leptonic
transitions; such decays can proceed via inner bremsstrahlung (IB), with a photon emitted by a charged particle in the initial or final state, via a
structure-dependent (SD) process, also known as direct emission (DE), which emits a photon from within the main transition, and the possible interference (INT)
between IB and SD.

$K^{\pm} \rightarrow \pi^{\pm} \pi^0 \gamma$ has a decay amplitude $W^2 = \frac{(p_\pi \cdot p_\gamma)(p_K \cdot p_\gamma)}{m_K^2 m_\pi^2}$; defining
$T_\pi^*$ as the $\pi^{\pm}$ kinetic energy and integrating it away from 
$\frac{d\Gamma^{\pm}}{dW} = \frac{d\Gamma_{IB}^{\pm}}{dW} [ 1 + 2 m_K^2 m_\pi^2 cos(\pm \phi + \delta_1^1 - \delta_0^1) X_E W^2 + m_K^4 m_\pi^4 (|X_E|^2 + |X_M|^2) W^4 ]$, it is possible to isolate the IB component, known from $K^{\pm} \rightarrow \pi^{\pm} \pi^0$ and QED corrections, and have the DE amplitude
containing electric (XE) and magnetic (XM) dipole terms. The INT component is interference between IB and electric DE (XE) amplitudes.
NA48/2 final results \cite{NA482PiPi0g} related to this decay are: Frac(DE)  = $( 3.32 \pm 0.15 \pm 0.14)  10^{-2}$, 
Frac(INT) = $(-2.35 \pm 0.35 \pm 0.39)  10^{-2}$ (first evidence) and 
$A_{CP} = \left|\frac{\Gamma^+ - \Gamma^-}{\Gamma^+ + \Gamma^-}\right| < 1.5 \times 10^{-3}$ (first measurement).
NA48/2 has also a preliminary result on $K^{\pm} \rightarrow \pi^{\pm} \pi^0 e^+ e^-$, which is mediated mainly by the process
$K^{\pm} \rightarrow \pi^{\pm} \pi^0 \gamma^* \rightarrow \pi^{\pm} \pi^0 e^+ e^-$ \cite{PiPi0ee}.
As for the previous decay, DE and INT amplitudes depend on XE and XM form factors, but its short distance contributions make it sensitive to New Physics (NP);
this result, which has $\approx 4500$ events in the signal region (2003+2004 data), with a small fraction of background from 
$K^{\pm} \rightarrow \pi^{\pm} \pi^0 \pi_D^0$
($\pi_D^{0} \rightarrow e^{+} e^- \gamma_{LOST}$) and $K^{\pm} \rightarrow \pi^{\pm} \pi_D^0$ ($\pi_D^{0} \rightarrow e^{+} e^-) + \gamma_{ACC}$ 
is also a first observation of this decay.

$K^{\pm} \rightarrow \pi^{\pm} \gamma \gamma$ is peculiar because parametrizing its branching fraction as a function of $z = \frac{m_{\gamma \gamma}^2}{m_K^2}$, 
its value depends on a single unknown $O(1)$ parameter \^c; a previous measurement performed at BNL E787, based on 31 candidates, gives 
$BR = (1.10 \pm 0.32)\times10^{-6}$ \cite{E787Pigg}.
The current measurement at NA62, based on $\approx 300$ event candidates with $O(10\%)$ background ($z>0.2$), gives two values for \^c, corresponding to
two different expansions in ChPT: \^c$ = 1.56 \pm 0.22_{stat} \pm 0.07_{syst} = 1.56 \pm 0.23$ for $O(p^4)$ and 
\^c$ = 2.00 \pm 0.24_{stat} \pm 0.09_{syst} = 2.00 \pm 0.26$ for $O(p^6)$. The sensitivity is insufficient to distinguish between the two models.
The model dependent result, not published yet, is $BR = (1.01 \pm 0.06)\times10^{-6}$.

$K \rightarrow e \nu_{e} \gamma$ SD+ has a differential amplitiude 
($\frac{d^2\Gamma_{SD}}{dxdy} = \frac{m_K^5 \alpha G_F^2 |V_{us}|^2}{64 \pi^2}$ $\left[ (F_V + F_A)^2 f_{SD+}(x,y) + (F_V - F_A)^2 f_{SD-}(x,y) \right]$)
with an explicit dependence on a vector and a axial FF; $f_{SD+}$, $f_{SD-}$, with $x = \frac{2E_\gamma^*}{m_K}$, $y = \frac{2E_e^*}{m_K}$,
represent known and different kinematics, being the + or - related to the polarization of the photon. A measurement by KLOE \cite{KLOEKe2g}, with 4\% accuracy,
is compatible with $O(p^4)$ FF (constant). NA62 has a preliminary result based on $\approx 10000$ event candidates.
\section{Lepton universality}
A powerful test of lepton universality has been performend by NA62 measuring the ratio 
$R_{K} = \frac{\Gamma(K \rightarrow e \nu_{e})}{\Gamma(K \rightarrow \mu \nu_{\mu})}$, where $BR(K \rightarrow e \nu) \approx O(10^{-5})$, being helicity
suppressed, and $BR(K \rightarrow \mu \nu) \approx 63\%$.
In the SM $R_{K} = \left(\frac{m_{e}}{m_{\mu}}\right)^{2}
\left(\frac{m_{K}^{2}-m_{e}^{2}}{m_{K}^{2}-m_{\mu}^{2}}\right)^{2}
(1+\delta R_{QED}) = (2.477 \pm 0.001)10^{-5}$; the advantages of this observable are the cancellation of hadronic uncertainties
in the ratio, an helicity suppression $\approx10^{-5}$ and small radiative correction (few \%) due to $K \rightarrow e \nu_{e} \gamma (IB)$,
by definition included into $R_{K}$. This results in a very clean theoretical prediction \cite{RK}. The experimental status in 2008 was
$R_{K} = (2.45 \pm 0.11)10^{-5}$ ('70s measurements), with $\delta R_{K} / R_{K} \approx 4.5\%$; KLOE in 2009 gives
$R_{K} = (2.493 \pm 0.031)10^{-5}$ \cite{KLOERK}, with $\delta R_{K} / R_{K} \approx 1.3\%$.

Given its small value within the SM, $R_{K}$ is potentially sensitive to NP; expected effects are within $\delta R_{K} / R_{K} \approx 10^{-4}-10^{-2}$.
A specific case of Minimal Supersymmetric Standard Model (MSSM) predicts 
$R_{K}^{MSSM} = R_{K}^{SM} \left[1+\left(\frac{m_{K}}{m_{H}}\right)^{4}\left(\frac{m_{\tau}}{m_{e}}\right)^{2}|\Delta_{13}|^{2} \tan^{6}\beta\right]$, which,
with $m_{H}=500$GeV/$c^2$,$|\Delta_{13}|=5\times10^{-4}$ and $\tan\beta=40$, gives  $R_{K}^{MSSM} = R_{K}^{SM} (1+0.013)$ \cite{MSSMRK1,MSSMRK2}.
$\pi$ and $B$ have the same effect, but in $R_{\pi}$ it's suppressed by $(m_{\pi}/m_{K})^4\approx10^{-3}$, $B \rightarrow e \nu_{e}$ is out of reach 
and $\frac{B \rightarrow \mu \nu_{\mu}}{B \rightarrow \tau \nu_{\tau}}$ has $\approx50\%$ enhancement.

NA62 final result (full data sample) on $R_{K}$ is based on 145,958 $K_{e2}$ candidates, a positron ID efficiency $(99.28 \pm 0.05)\%$ and
a background $B/(S+B) = (10.95 \pm 0.27)\%$; a fit was performed over 40 statistically independent measurements (4 data samples $\times$ 10 momentum bins),
with $\chi^2/ndf =47/39$, giving $R_{K}=(2.488 \pm 0.007_{stat} \pm 0.007_{syst})\times10^{-5}$
\begin{table}[!htbp]
\begin{center}
\begin{minipage}[!htbp]{0.45\textwidth}
        \centering
\begin{tabular}{l r}
        \hline
        Source                           					&      $\delta R_{K}\times10^{5}$\\
        \hline
         Statistical                     					&       0.007\\
         $K \rightarrow \mu \nu_\mu$    					&       0.004\\
         $K \rightarrow e \nu_{e} \gamma$ ($SD^{+}$)    			&       0.002\\
         $K \rightarrow \pi^0 e \nu_{e}$, $K \rightarrow \pi \pi^0$       	&       0.003\\
         Beam halo                       					&       0.002\\
         Matter composition              					&       0.003\\
         Acceptance                      					&       0.002\\
         Positron ID                     					&       0.001\\
         DCH alignmnent                  					&       0.001\\
         1-track trigger                 					&       0.001\\
        \hline
         Total                           					&       0.010\\
        \hline
\end{tabular}
        \caption{$R_{K}$ uncertainties}
        \label{tab:RKunc}
\end{minipage}
\begin{minipage}[!htbp]{0.45\textwidth}
        \centering
        \includegraphics[width=\textwidth]{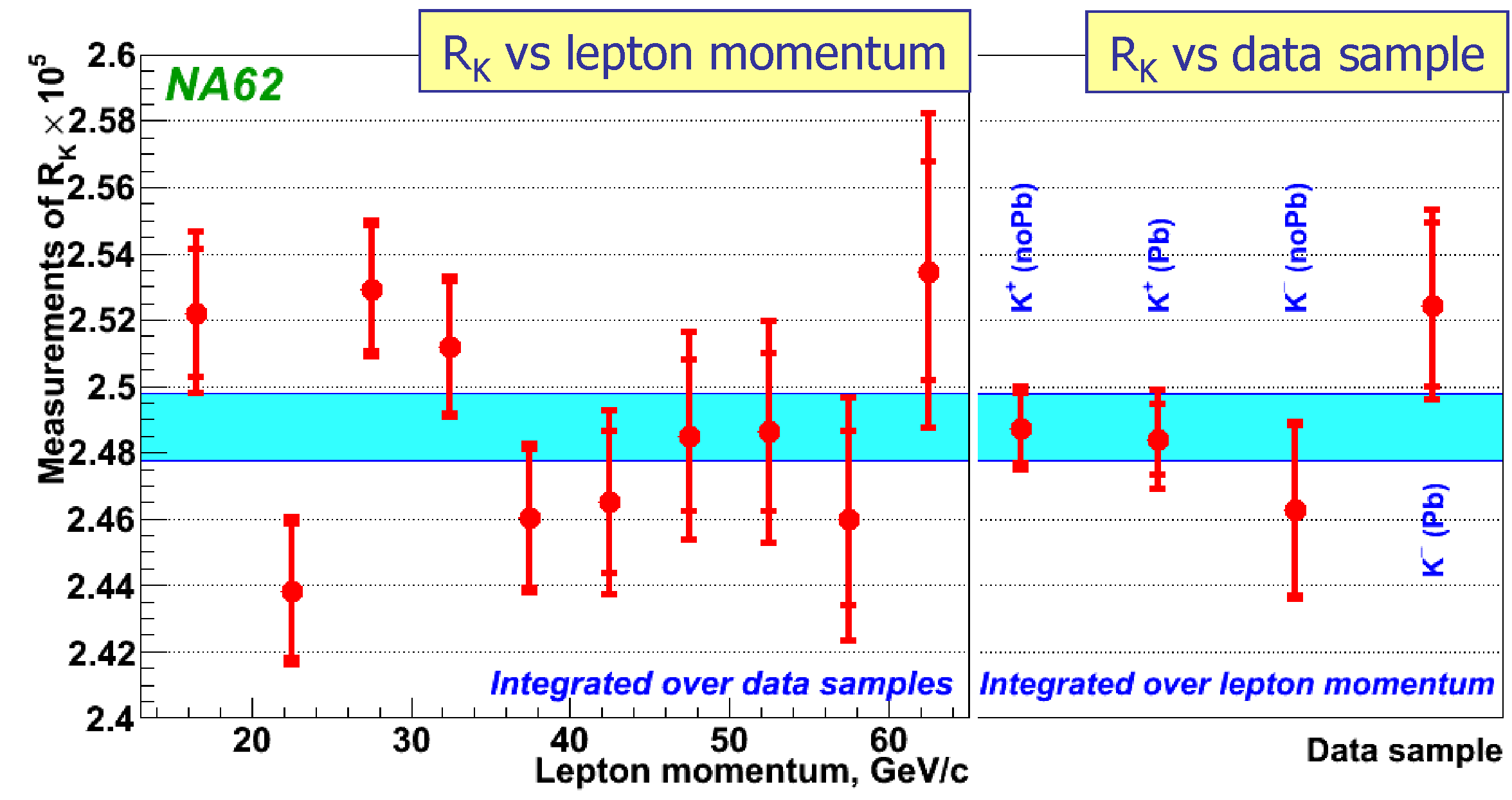}
        \caption{$R_{K}$ independent measurements}
        \label{fig:RkFinalFull}
\end{minipage}
\end{center}
\end{table}

$R_{K}$ world average changed from $(2.493 \pm 0.025)\times10^{-5}$ in 2010 to $(2.488 \pm 0.009)\times10^{-5}$ in 2012, going from 1.0\% to 0.36\% precision.
%
\section{Flavour Changing Neutral Currents}
Processes which involve Flavour Changing Neutral Currents (FCNC) in the SM are suppressed by the Glashow-Iliopoulos-Maiani (GIM) mechanism \cite{GIM}, 
therefore they are in principle sensitive to NP effects. NA48/2 has results also for $K^{\pm} \rightarrow \pi^{\pm} l^+ l^-$, which is an example of this
category. It is a loop induced process ($K^{\pm} \rightarrow \pi^{\pm} \gamma^*$) expected to have a $BR \approx 10^{-7}$. For
$K^{\pm} \rightarrow \pi^{\pm} e^+ e^-$ \cite{NA482Piee} NA48/2 has $\approx 7200$ event candidates ($>4\times$ world statistics), $<1\%$ background
and quotes $BR = (3.11 \pm 0.12) \times 10^{-7}$, with $A_{CP} < 2.1 \times 10^{-2}$. For $K^{\pm} \rightarrow \pi^{\pm} \mu^+ \mu^-$ \cite{NA482Pimumu}
it has $\approx 3100$ event candidates, $(3.3 \pm 0.7)\%$ background, and quotes $BR = (9.62 \pm 0.25) \times 10^{-8}$, with $A_{CP} < 2.9 \times 10^{-2}$.
\subsection{Golden modes}
$\kpnnn$, in its charged and neutral kaon variations, is among the few golden decays; their theoretical prediction exploits the factorization of the
matrix element, allowing to extract the hadronic component from the experimental value of $BR(K \rightarrow \pi e \nu)$. This results in
$BR_{SM}(\klnn) = (2.43 \pm 0.39 \pm 0.06)10^{-11}$, and $BR_{SM}(\kpnn) = (7.81 \pm 0.75 \pm 0.29)10^{-11}$, where the uncertainities are related to CKM
matrix elements knowledge and theory respectively.
The only existing measurement, related to $\kpnn$, is based on 7 events (E787/949): \alert{$(1.73^{+1.15}_{-1.05})10^{-10}$}. A new measurement at the level
of accuracy of the theoretical prediction would lead to an independent determination of $V_{td}$ with $\approx 7\%$ precision, and open a new set of
tests of the SM against NP scenarios.
Several experiments are foreseen in the near future to measure $\kpnnn$ decays.
\begin{tabular}{l l c c c c}
	\hline
	Expt 		& Primary beam			& Intensity      		&   SM  	& Start date		 & Total\\
	     		&             			&   (ppp)       		& evts/yr	& + run yrs		 & SM evts\\
	\hline
	NA62 		& SPS 450 GeV			& $3 \pm 10^{12}$		& 55		& 2014+2		& 110\\
	FNAL $K^{\pm}$	& Project X 8 GeV		& $2 \pm 10^{14}$		& 250		& 2018+5		& 1250\\
	ORKA		& Tevatron up $<$150 GeV	& $5 \pm 10^{13}$		& 120		& 2018+5		& 600\\
	E14(KoTO)	& JPARC-I 30 GeV		& $2 \pm 10^{14}$		& 1-2		& 2013+3		& 3-7\\
	E14		& JPARC-II 30 GeV		& $3 \pm 10^{14}$		& 30		& 2020+3?		& 100\\
	FNAL KL 	& Booster 8 GeV			& $2 \pm 10^{13}$		& 30		& 2016+2		& 60\\
	FNAL KL		& Project X 8 GeV		& $2 \pm 10^{14}$		& 300		& 2018+5		& 1500\\
	\hline
\end{tabular}

\subsection{Measurement of $BR(\kpnn)$ at NA62}
\begin{figure}
        \includegraphics[width=\textwidth,height=0.3\textwidth]{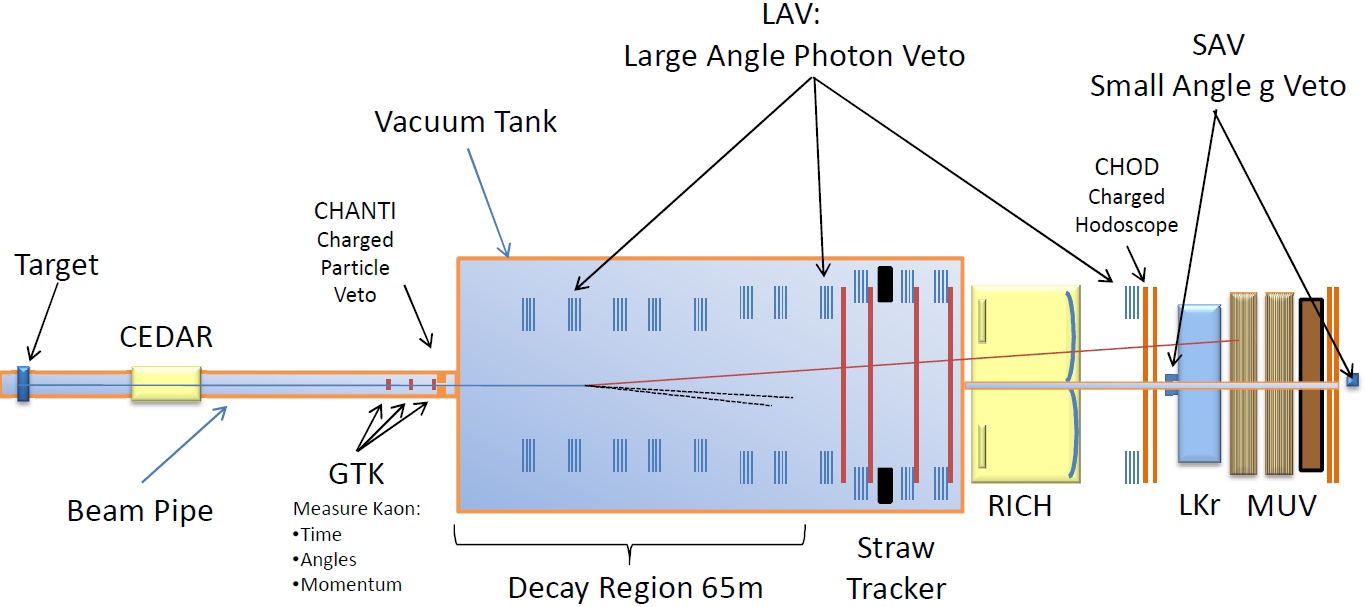}
        \caption{NA62 layout}
        \label{fig:NA62Layout}
\end{figure}

An example of this new generation of experiments is NA62 at CERN.
NA62 aims to a measurement at 10\% ($\approx$ SM prediction accuracy); it is foreseen to collect 100 SM events in 2 years data taking. The background rejection
necessary to achieve the needed sensitivity is obtained expointing the kinematics of kaon decays combined with particle identification and photon vetoes. 
The unseparated charged hadron beam, composed by $p/\pi/K$ (positron free, $K\approx 6\%$, $p\approx 23\%$), at 75 GeV/$c$ ($\Delta P/P \approx 1\%$)
will provide $4.8\times10^{12}$ kaon decays/year, corresponding to an integrated beam rate of 750 MHz.
The layout of the experiment is shown in fig. \ref{fig:NA62Layout} and it makes use of state of the art detectors for new precision frontier down to $10^{-12}$.
Technical run in 2012 and physics data taking in 2014-2016

\section{Exotic}
Among the exotic channels a recent result from NA48/2 is the search for $K^{\pm} \rightarrow \pi^{\pm} \mu^+ \mu^+$, a lepton number violating process
which is searched looking for the wrong-sign events in $K^{\pm} \rightarrow \pi^{\pm} \mu^+ \mu^-$ data; no evidence was found and NA48/2 quotes
$BR(K^{\pm} \rightarrow \pi^{\mp} \mu^\pm \mu^\pm) < 1.1 \times 10^{-9}$  (90\% CL), which is 3 times better than the previous limit from E865 \cite{exotic}.

\section{Summary}
Kaon physics continues to be a good tool for investigation in the flavour sector. Chiral Perturbation Theory and experimental determination of form factors 
provide a constantly improving tool for future precision measurements. All measurements are currently in agreement with the SM. A new generation of experiments 
is starting to explore ultra rare decays, opening a new chapter of tests for the SM and precision measurements previously not accessible: NA62 and KoTO are in 
construction and will start taking data in the next two years; these detectors will be able to improve current measurements

\end{document}